\documentclass[12pt]{article}
\usepackage{amssymb}
\usepackage{latexsym}
\usepackage{amsmath}
\usepackage{graphicx}

\def\vp{\varphi}
\def\al{\alpha}
\title{\LARGE{Slowly rotating Einstein-Maxwell-dilaton black hole and some aspects of its thermodynamics }}
\author{M. M. Stetsko\footnote{E-mail: mstetsko@gmail.com}\
\\
  {\small Department for Theoretical Physics, Ivan Franko National University of Lviv,}\\
{\small 12 Drahomanov Str., Lviv, UA-79005, Ukraine
         }}

\begin{document}
\maketitle

\abstract{A slowly rotating black hole solution in Einstein-Maxwell-dilaton gravity was considered. Having used the obtained solution we investigated thermodynamic functions such as black hole's temperature, entropy and heat capacity. In addition to examine thermodynamic properties of the black hole extended technique was applied. The equation of state of Van der Waals type was obtained and investigated. It has been shown that the given system has phase transitions of the first as well as of the zeroth order for the temperatures below a critical one which is notable feature of the black hole. A coexistence  relation for two phases was also considered and latent heat was calculated. In the end, critical exponents were calculated.}

\section{Introduction}
Thermodynamics of black holes has been investigated intensively for over four decades. This area of research is considered to bring some links between general relativity and quantum mechanics and might be very useful for verification of different approaches to quantum gravity \cite{Carlip_IJMPD14}. It is supposed that all the approaches in the quasiclassical limit should lead to the same thermodynamic behaviour of black holes and to be in agreement with the laws of black hole mechanics \cite{Bardeen_CMP73} which were developed without any knowledge about underlying theory of quantum gravity. It might be treated as a theory that allows to correct or even get rid of some deficit approaches to quantum gravity. But it should be noted that such different approaches to quantum gravity as String Theory \cite{Strominger_PLB96} and Loop Quantum Gravity \cite{Rovelli_PRL96} give rise to the same expression for black hole entropy-area relation given by celebrated Bekenstein-Hawking formula. 

New great interest in thermodynamics of black holes has been observed for the latter decade. This revival of activity can be explained by the fact that in recent works it was proposed to extend the complex of thermodynamic variables which had been used in black holes' theory and consider the cosmological constant as one of those thermodynamic values \cite{Kastor_CQG09,Cvetic_PRD11,Dolan_CQG11,Dolan_PRD11,Kubiznak_JHEP12, Kubiznak_JHEP12_2}. It was suggested that for a gravitational system for example for a black hole such a standard notion of thermodynamics as pressure can be introduced and it is proportional to the cosmological constant. In case of Reissner-Nordstrom-AdS (RN-AdS) black hole one has a simple relation between the thermodynamic pressure and the cosmological constant: $P=-\Lambda/8\pi$ \cite{Kastor_CQG09,Cvetic_PRD11,Dolan_CQG11,Dolan_PRD11}. Since the cosmological constant here is supposed to be a variable it changes the identification of black hole mass in thermodynamics. In contrast to the well-established treatment of the mass as the black hole's internal energy in the extended approach it is considered as thermodynamic enthalpy, namely: $M=H=U+PV$. The second point that is strongly related to the introduced notion of the pressure is the definition of its thermodynamically conjugate value, namely it is the volume which can be introduced by the derivative: $V=(\partial H/\partial P)_S$. The extended (enlarged) thermodynamic phase space allowed to examine phase behaviour of black holes, namely it was shown that charged black hole possesses a phase transition that is completely analogous to Van der Waals liquid-gas phase transition and below the critical point it has the phase transition of the first order \cite{Kubiznak_JHEP12}. It should be noted that richer thermodynamic behaviour appears in case of more general setting, namely for Born-Infeld theory or for black holes in a space of higher dimensions. For example, reentrant phase transition appears \cite{Kubiznak_JHEP12_2,Altamirano_PRD13} or tricritical (triple) point was observed \cite{Altamirano_CQG13,Wei_PRD14}. The mentioned facts stimulated the deep interest and prolific investigations in this area of research which obtained the special name -- black hole chemistry and this name reflects the similarity in thermodynamic behaviour of black holes and ordinary condensed matter systems such as gases, liquids, multicomponent fluid systems and so on. The overview of recent developments and possible future directions of this subfield of black hole thermodynamics is given in \cite{Kubiznak_CQG2017}.

General Relativity is very successful theory in explanation of different phenomena on the scales from the planetary one up to cosmological values.  But nonetheless, different approaches to quantum gravity, namely the String Theory are supposed to modify the gravitational Einstein's term replacing it with a sum of ones which takes into account higher orders corrections of curvature. In the low energy limit of the String Theory standard Einstein's term plus nonminimally coupled dilaton field can be obtained \cite{Witten_87}. It is worth stressing that dilaton field minimally coupled to Einstein's gravity appears as a result of a dimensional reduction of the standard General Relativity in higher dimensions onto a space of a lower dimension. So, such low energy limit of the String Theory can be considered as Einstein gravity with additional dilaton field. This model has attracted considerable attention for latter two decades, namely different types of black holes/strings solutions were investigated \cite{Gibbons_NPB88,Garfinkle_PRD91,Witten_PRD91,Gregory_PRD93,
Rakhmanov_PRD94,Poletti_PRD94,Chan_NPB95,Cai_PRD96,Cai_PRD98,
Clement_PRD03,Cai_PRD04,Gao_PRD04,Gao_PLB05,Yazadjiev_CQG05,Astefanesei_PRD06,Mann_JHEP06,
Charmousis_PRD09,Gouteraux_JHEP11,Gouteraux_JHEP12,Quevedo_PRD16}. We point out that in bigger part of those works static solutions were studied, whereas the rotating solutions were examined for some particular choice of dilaton coupling parameter \cite{Kunduri_PLB05,Yazadjiev_PRD05,Kunz_PLB06,Brihaye_CQG07,Charmousis_JHEP07,Ghosh_PRD07,
Blazquez_PRD14,Knoll_PRD16,Kleihaus_Entropy16}.  It should be noted that the special interest has been paid to the investigation of Einstein-dilaton black holes with additional dilaton potential of a so called Liouville-type form \cite{Poletti_PRD94,Chan_NPB95, Cai_PRD98,Gao_PRD04,Yazadjiev_CQG05,Sheykhi_PRD06,Sheykhi_IJMPA07,Sheykhi_PRD07,
Sheykhi_PRD08,Sheykhi_PRD08,Sheykhi_PLB08,Sheykhi_PRD08_02,Sheykhi_GRG10}. The Liouville-type potential allows to introduce terms which can be treated as a generalization of the known cosmological constant term. This fact can be used for examination of possible extensions of AdS-CFT correspondence. It was also supposed that linear dilaton spacetime which appears as a near horizon limit of a dilatonic black hole might possess some holographic properties \cite{Aharony_JHEP98}. From the other side new black hole solutions which are not asymptotically flat nor of AdS-type (or dS-type) might be a good laboratory for application or testing of methods developed for the mentioned above types of black hole solutions. The interest to the dilatonic black holes with Liouville dilaton potential was quite deep, namely black hole solutions with different types of nonlinear electromagnetic field coupled to dilaton filed were obtained \cite{Sheykhi_PRD07_3,Sheykhi_IJMPD08,Sheykhi_PRD14,Sheykhi_PRD14_2,
Kord_PRD15_0,Hendi_PLB17}. Considerable attention was paid to the investigation of thermodynamic properties of corresponding dilatonic black holes. It should be noted that the standard approach as well as the extended technique were applied for the study of their thermodynamics \cite{Sheykhi_PRD07,Sheykhi_IJMPD08,Kord_PRD15_0,Sheykhi_PRD10,
Deghani_2015,Kord_PRD15,Hendi_PRD15,Kord_JHEP16,Zhao_AHEP16,
Mo_PRD16,H_F_Li_2016,Kord_EPJC17,Dayyani_PRD17,Dehyadegari_2017}.

In our work we examine a slowly rotating Einstein-Maxwell-dilaton black hole with linear Maxwell field and electromagnetic-dilaton coupling. We also take into account a dilaton potential of the Liouville form which consists of two terms, in particular one of them is supposed to be a kind of generalization of a cosmological constant term. We also consider thermodynamics of given black hole solution. For the first, assuming that corresponding cosmological constant is fixed we obtain temperature and heat capacity. Secondly, using the modern technique where thermodynamic pressure is introduced we obtain the equation of state and derive Gibbs potential and investigate critical behaviour.

This work is organised in the following way. In the second section the slowly rotating black hole is obtained and examined. The third section is devoted to thermodynamics of the black hole, namely relations for the temperature and heat capacity are investigated. In the forth section the equation of state and Gibbs potential for the black hole are obtained and investigated. The fifth section is devoted to the derivation of critical exponents. Finally, the last  section  contains some conclusions.    

\section{Field equations and their solution}
We consider $n+1$-dimensional gravity ($n\geqslant 3$) with dilaton and linear Maxwell  fields. The action integral for this system can be written  in the form:
\begin{equation}\label{action_int}
S=\frac{1}{16\pi}\int_{{\cal M}} {\rm d}^{n+1}x\sqrt{-g}\left(R-\frac{4}{n-1}\nabla^{\mu}\Phi\nabla_{\mu}\Phi-V(\Phi)-e^{-4\alpha\Phi/(n-1)}F_{\mu\nu}F^{\mu\nu}\right)-\frac{1}{8\pi}\int_{\partial{\cal M}}{\rm d}^{n}x\sqrt{-\gamma}\Theta(\gamma)
\end{equation}
where $R$ is the scalar curvature, $\Phi$ is the dilaton field, $V(\Phi)$ denotes the potential which depends on the dilaton field, $\al$  denotes the dilaton-electromagnetic coupling parameter and  $F_{\mu\nu}$ is the electromagnetic field tensor which is defined in the standard way, namely $F_{\mu\nu}=\partial_{\mu}A_{\nu}-\partial_{\nu}A_{\mu}$  and here $A_{\mu}$ is a component of the electromagnetic potential. The second term in the action (\ref{action_int}) corresponds to the so called Gibbons-Hawking boundary term which makes the variation of the action well defined and allows us to obtain conserved quantities. The $\partial {\cal M}$ denotes the boundary of the manifold ${\cal M}$, $\gamma_{ab}$ is the metric on the boundary and $\Theta $ is the trace of extrinsic curvature tensor $\Theta_{ab}$ on the boundary. 

Varying the action integral (\ref{action_int}) with respect to the gravitational field, represented by the metric tensor $g_{\mu\nu}$, dilaton field $\Phi$ and the electromagnetic potential $A_{\mu}$ one obtains field equations which can be written in the following form:
\begin{eqnarray}
R_{\mu\nu}=\frac{g_{\mu\nu}}{n-1}\left(V(\Phi)-e^{-4\al\Phi/(n-1)}F_{\rho\sigma}F^{\rho\sigma}\right)+\frac{4}{n-1}\partial_{\mu}\Phi\partial_{\nu}\Phi+2e^{-4\al\Phi/(n-1)}
F_{\mu\sigma}F_{\nu}^{\sigma};\label{einstein}\\
\nabla_{\mu}\nabla^{\mu}\Phi=\frac{n-1}{8}\frac{\partial V}{\partial \Phi}-\frac{\al}{2}e^{-4\al\Phi/(n-1)}F_{\rho\sigma}F^{\rho\sigma};\\
\nabla_{\mu}(e^{-4\al\Phi/(n-1)}F^{\mu\nu})=0\label{em_equation}
\end{eqnarray}
We are interested in the solution of the field equations which represents a slowly rotating  black hole so we should make some  choice of the supposed form of the metric. To find this solution we represent the metric in the form:
\begin{equation}\label{metric}
ds^2=-W(r)dt^2+\frac{dr^2}{W(r)}-2af(r)\sin^2{\theta}dtd\vp+r^2R^2(r)(d\theta^2+\sin^2{\theta}d\vp^2+\cos^2{\theta}d\Omega^2_{n-3})
\end{equation}
and here the metric functions $W(r)$, $f(r)$ and $R(r)$ are supposed to depend on the radial coordinate $r$ only and parameter $a$ is related to the angular momentum (or angular velocity) of the black hole. It is worth noting that similar problem was considered in  papers \cite{Sheykhi_PRD08_02} but here we make a bit different choice of the dilaton potential $V(\Phi)$. It should also be pointed out that higher dimensional Einstein-Maxwell-dilaton black hole with multiple rotational parameters was considered in the work \cite{Sheykhi_GRG10}.  As it was pointed out in the work \cite{Sheykhi_PRD08} the only term in the metric that appears due to slow rotation is the $g_{t\vp}$ term which is of the order of ${\cal O}(a)$ and this fact motivates the choice of the third term in the metric (\ref{metric}). Infinitesimal rotation does not affect on the dilaton field $\Phi$ and the electromagnetic potential acquires $A_{\vp}$ term which is also of the order if ${\cal O}(a)$. In the limit $a\rightarrow 0$ the static solution is recovered.

From the equation (\ref{em_equation})  one can easily obtain the components of the gauge field tensor. The leading term of the electromagnetic field tensor takes the form:
\begin{equation}
F_{tr}=\frac{qe^{4\al\Phi/(n-1)}}{r^{n-1}R^{n-1}}
\end{equation}
and here the constant of integration $q$ is related to the electric charge of the black hole which will be calculated a bit later. As it was noted before, even the slow rotation we consider adds an ``angular\rq\rq{} component $A_{\vp}$ to the electromagnetic field potential, which can be chosen in the form \cite{Sheykhi_PRD08}:
\begin{equation}
A_{\vp}=aqh(r)\sin^2{\theta}
\end{equation}
and $h(r)$ is a function of the radial coordinate and evident form of this function can be obtained from the field equations (\ref{einstein})-(\ref{em_equation}). Other components of the electromagnetic field tensor would be proportional  to the $a$ and as a result the term $F_{\mu\nu}F^{\mu\nu}$ would be proportional to the parameter $a^2$ and we do not take these terms into consideration in the field equations (\ref{einstein})-(\ref{em_equation}).

The dilaton potential $V(\phi)$ is taken in the so called Liouville form:
\begin{equation}\label{dilat_potential}
V(\Phi)=\Lambda_0 e^{\lambda_0\Phi}+\Lambda e^{\lambda\Phi}.
\end{equation}

To obtain a solution of the field equations (\ref{einstein})-(\ref{em_equation}) the following ansatz for the function $R(r)$ can be used \cite{Sheykhi_PRD07,Sheykhi_PRD08}:
\begin{equation}\label{ansatz}
R(r)=e^{2\al\Phi/(n-1)}.
\end{equation} 

Having used the latter ansatz we can solve the field equations (\ref{einstein})-(\ref{em_equation}) and obtain  the metric functions $W(r)$, $f(r)$, the dilaton field  $\Phi(r)$ and the function $h(r)$:
\begin{eqnarray}\label{func_w}
\nonumber W(r)=-mr^{1+(1-n)(1-\gamma)}+\frac{(n-2)(1+\al^2)^2}{(1-\al^2)(\al^2+n-2)}b^{-2\gamma}r^{2\gamma}-\\\frac{\Lambda(1+\al^2)^2}{(n-1)(n-\al^2)}b^{2\gamma}r^{2(1-\gamma)}+\frac{2q^2(1+\al^2)^2}{(\al^2+n-2)(n-1)}b^{2(2-n)\gamma}r^{-2(n-2)(1-\gamma)},\label{metric_W}
\end{eqnarray}
\begin{eqnarray}\label{func_f}
\nonumber f(r)=\frac{m(\al^2+n-2)}{(n-2)(1+\al^2)}b^{(n-1)\gamma}r^{1+(1-n)(1-\gamma)}+\frac{\Lambda(1+\alpha^2)}{(n-1)(n-2)}\times\\\frac{(n+\al^2-2)}{(n-\al^2)}b^{(n-1)\gamma}r^{2(1-\gamma)}-\frac{2q^2(1+\al^2)}{(n-1)(n-2)}b^{(1-n)\gamma}r^{2(2-n)(1-\gamma)},
\end{eqnarray}
\begin{eqnarray}\label{func_phi}
\Phi(r)=\frac{\al(n-1)}{2(1+\al^2)}\ln{\left(\frac{b}{r}\right)},\\
h(r)=\frac{1}{n-2}r^{-1+(3-n)(1-\gamma)},
\end{eqnarray}
where the parameter $m$ is an integration constant related to black hole\rq{}s mass (as it will be shown below) and $\gamma=\al^2/(1+\al^2)$. We should pay special attention to the parameter $b$ which appears in the written above relations (\ref{func_w})-(\ref{func_phi}) which is also an integration constant but it does not have direct physical meaning, looking at the relation (\ref{func_phi})  we can treat it as a ``scaling'' parameter for the dilation field, but the physical meaning of this parameter is unclear. Since we have just assumed that the parameter $b$ is a ``scaling'' one, we might perform a coordinate transformation, namely $\bar{r}=r/b$ to simplify written above relations (\ref{func_w})-(\ref{func_phi}), but it is easy to check that such rescaling does not simplify the mentioned relations and possibly requires the corresponding rescaling of the mass $m$ and charge $q$ parameters and of the parameter $\Lambda$ what is not convenient and has no advantages. Since the parameter $b$ is arbitrary we can set it equal to unity ($b=1$) for simplification, but in what follows we keep it as it is in order to make our relations easily comparable with the previously obtained results, namely in the work \cite{Sheykhi_PRD08}. The obtained functions coincide with corresponding functions given in the paper \cite{Sheykhi_PRD08} when the parameter $\Lambda$ is set to zero ($\Lambda=0$).  It is worth stressing that parameters $\lambda_0$, $\lambda$ and $\Lambda_0$ are not arbitrary, whereas there is no restriction on the parameter $\Lambda$, which can be treated as a cosmological constant. To fully obey the  system of equations (\ref{einstein})-(\ref{em_equation}) the mentioned above parameters should be taken in the following form:
\begin{equation}
\Lambda_0=\frac{\alpha^2(n-1)(n-2)}{b^2(\al^2-1)},\quad \lambda_0=\frac{4}{\al(n-1)},\quad \lambda=\frac{4\al}{(n-1)}.
\end{equation}
 The relation (\ref{metric_W}) describes a multitude of causal structures of the dilatonic black hole. The obtained metric functions (\ref{func_w})-(\ref{func_phi}) show that in general the solution is not asymptotically flat nor (anti)-de Sitterian. It can be shown that the Kretschmann scalar $R_{\mu\nu\lambda\sigma}R^{\mu\nu\lambda\sigma}$ diverges only when $r\rightarrow 0$ and is finite for $r\neq 0$. It leads to the conclusion that the only physical singularity of the metric is located at the point $r=0$. We remark that the black hole solution is not well-defined when $\al=1$ this is so called string singularity and there is another singularity of the metric when $\al=\sqrt{n}$. Due to similarity of the obtained metric function (\ref{func_w}) with corresponding metric function obtained in the work \cite{Sheykhi_PRD07} the analysis of the causal structure would be the same as it was performed in the mentioned above publication.
 
In the limit when the dilaton parameter $\al$ is set to zero from the written above equations for the metric functions $W(r)$ and $f(r)$ it follows that:
\begin{eqnarray}
W(r)=1-mr^{2-n}-\frac{\Lambda}{n(n-1)}r^2+\frac{2q^2}{(n-1)(n-2)}r^{2(2-n)}\\
f(r)=mr^{2-n}+\frac{\Lambda}{n(n-1)}r^2-\frac{2q^2}{(n-1)(n-2)}r^{2(2-n)}
\end{eqnarray}
So, we obtain the metric functions for a (n+1)-dimensional slowly rotating charged Kerr-AdS black hole \cite{Aliev_PRD06}.

To obtain the black hole's mass we use the quasilocal concept \cite{Brown_PRD93}. Following the mentioned work of Brown and York \cite{Brown_PRD93} to derive quasilocal mass (or other quasilocal quantities) one should use ADM decomposition of the spacetime manifold ${\cal M}$ into evolving in time spacelike hypresurface $\Sigma_t$ times the real time interval $\mathbb{R}$ (${\cal M}=\Sigma_t\times\mathbb{R}$). Since our spacetime ${\cal M}$ has its boundary $\partial{\cal M}$, the evolving hypersurface $\Sigma_t$ also has its boundary which we denote as ${\cal B}$ and this boundary play important role in calculation of quasilocal quantities. Namely, according to the quasilocal concept to derive the mass of the black hole one should use the following relation:
\begin{equation}
M=\frac{1}{8\pi}\int_{{\cal B}}d^{n-1}\chi\sqrt{\sigma} N(k-k_0),
\end{equation}
we point out here that the latter integral is taken over the boundary ${\cal B}$ and $k$ is a trace of extrinsic curvature of the hypersurface ${\cal B}$ in embedding manifold $\Sigma_t$, $k_0$ denotes the trace of extrinsic curvature of a reference (background) metric, $\sigma$ is the determinant of the metric on the spacelike hypersurface ${\cal B}$ and $N$ denotes the lapse function in the mentioned above ADM decomposition of the spacetime metric.
As a result the black hole\rq{}s mass can be represented in the following form:
\begin{equation}\label{mass_bh}
M=\frac{(n-1)b^{(n-1)\gamma}\omega_{n-1}}{16\pi(1+\al^2)}m,
\end{equation}
where $\omega_{n-1}$ is the surface area of a $n-1$--dimensional unit hypersphere. We point here that since we consider non-asymptotically flat space-time the concept of quasi-local energy might be applied here to calculate black hole's mass and angular momentum and the use of the quasi-local concept allows  to obtain the first law of thermodynamic similarly as it was performed for asymptotically flat black holes \cite{Hawking_CQG96,Clement_PRD04}. It should be noted that the obtained  above relation for the  black hole's mass (\ref{mass_bh}) completely coincides with the corresponding relation calculated with help of the Abbott-Deser method \cite{Abbott_NPB82}. We also point out that the notion of quasi-local energy (and mass) is not unique, there are several various notions of it and any of them is useful for some particular problems of gravity and might have serious drawbacks when one tries to analyze some different problems \cite{Szabados_LRR04}.  

The electric charge of the black hole can be calculated via the relation (Gauss law):
\begin{equation}
Q=\frac{1}{4\pi}\int \exp\{-4\al\Phi/(n-1)\}*F
\end{equation} 
and here $*F$ denotes the form dual to the electromagnetic field form. It should be noted that the integration is performed over a closed spacelike hypresurface enclosing the black hole. Having used the latter relation we get:
\begin{equation}\label{charge}
Q=\frac{\omega_{n-1}}{4\pi}q
\end{equation}

\begin{figure}
\centerline{\includegraphics[scale=0.29,clip]{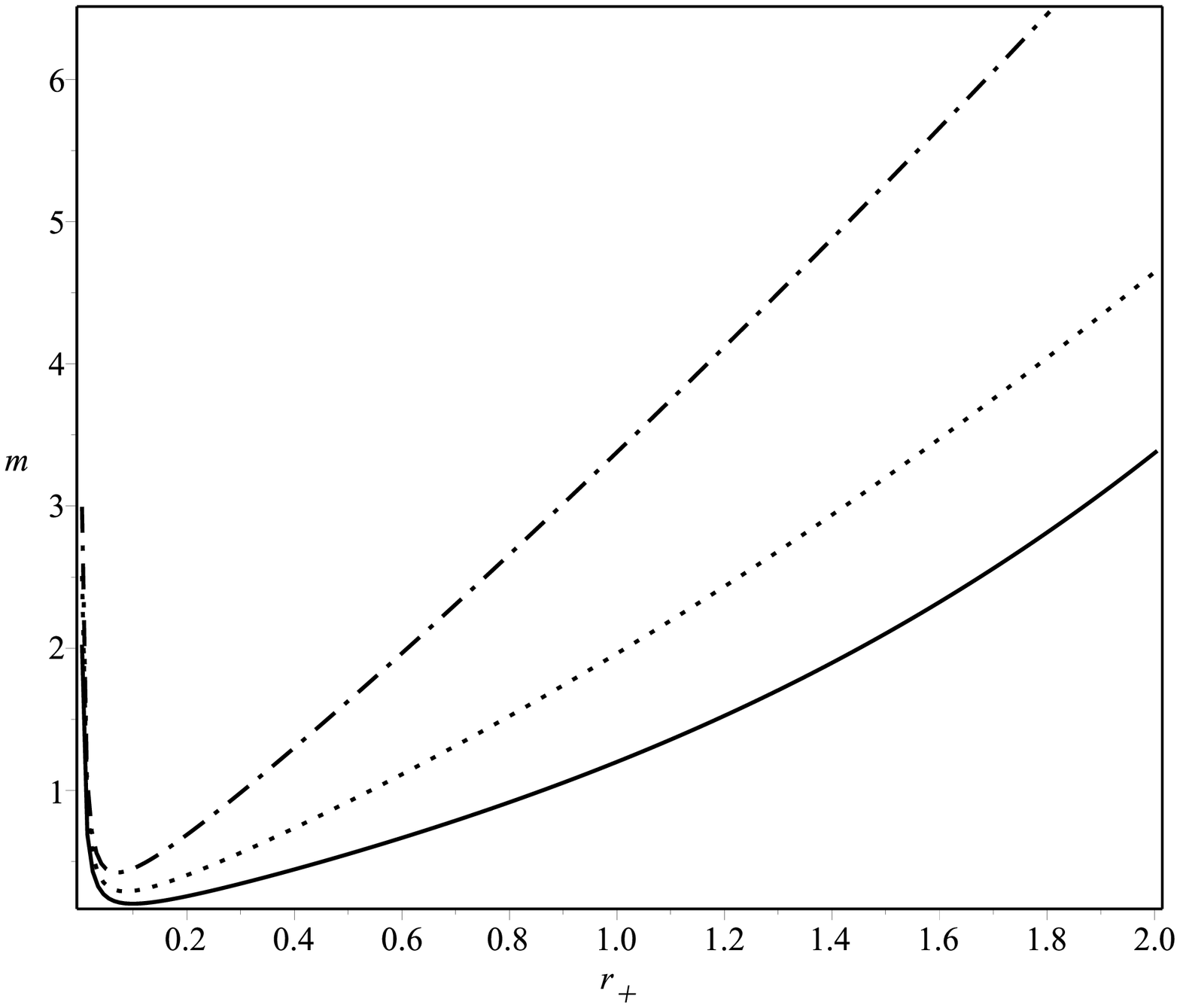}\includegraphics[scale=0.288,clip]{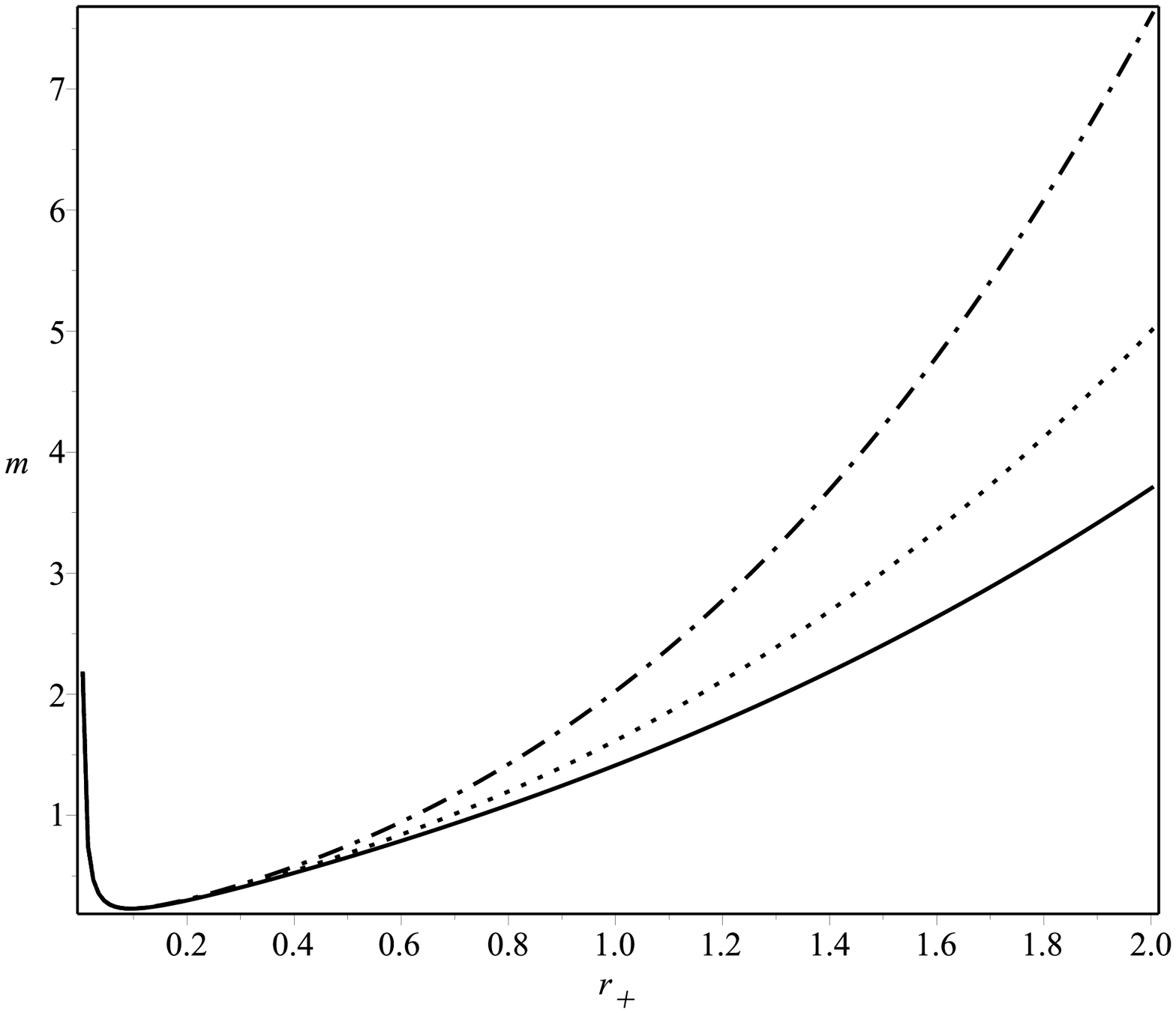}}
\caption{Mass parameter $m$ as a function of horizon radius $r_+$ for several values of parameter $\al$ (the left graph) or cosmological constant $\Lambda$ (the right one).  Namely, for the left graph the correspondence is the following: the solid curve corresponds to $\al=0.1$, the dotted curve corresponds to $\al=0.5$ and the dash-dotted on to $\al=0.7$, for all the curves here $\Lambda=-1$.  On the right graph the correspondence is as follows: solid, dotted and dash-dotted curves correspond to $\Lambda=-1$, $\Lambda=-2$ and $\Lambda=-3$ respectively, for all the lines $\al=0.3$. All the other parameters are held fixed for both graphs  ($n=3$, $b=1$, $q=0.1$).}\label{mass_gr1}
\end{figure}

The mass parameter can be rewritten in terms of the radius of black hole\rq{}s horizon:
\begin{eqnarray}\label{mass_param}
\nonumber m(r_+)=(1+\al^2)^2r_+^{(n-1)(1-\gamma)-1}\left(\frac{(n-2)}{(1-\al^2)(\al^2+n-2)}b^{-2\gamma}r_+^{2\gamma}\right.\\-\left.\frac{\Lambda}{(n-1)(n-\al^2)}b^{2\gamma}r_+^{2(1-\gamma)}+\frac{2q^2}{(\al^2+n-2)(n-1)}b^{2(2-n)\gamma}r_+^{2(2-n)(1-\gamma)}\right). 
\end{eqnarray}
The behaviour of the mass parameter as a function of the horizon radius is represented on the Fig. [\ref{mass_gr1}].  All the plots show that the mass parameter and as a consequence the black hole's mass has a point of minimal mass which corresponds to some specific value of horizon $r_+$. Decrease or increase of the horizon radius leads to the the increase of the mass. This fact is not specific peculiarity of the dilatonic black holes, the similar situation takes place for a charged black hole in anti-de Sitter universe. The nonmonotonic behaviour of the mass can be explained by the competition between the terms corresponding to electromagnetic and nonelectromagnetic parts. For small $r_+$ the electromagnetic part becomes dominant which leads to the decreasing of the mass with the decrease of the horizon radius $r_+$, whereas for large values of $r_+$ the other term becomes leading which gives the increasing of the mass when the horizon radius goes up.    

Since the black hole we consider is a rotating one it is important to calculate angular momentum for it. To find the angular momentum quasilocal formalism can be used again \cite{Brown_PRD93}. It should be noted that the mentioned approach works for Einsteinian general relativity with arbitrary matter field which should be nonderivatively coupled to gravity. The angular momentum is supposed to be a conserved quantity and its value can be obtained from the information about the behaviour of gravitational field on the boundary of a $n$-dimensional hypersurface evolving in time. Namely this information is encoded in the boundary term of the action (\ref{action_int}). Similarly as it was done in \cite{Sheykhi_PRD08} one can write the boundary stress-energy tensor:
\begin{equation}\label{stress-energy}
T_{ab}=\Theta_{ab}-\gamma_{ab}\Theta
\end{equation}
which results from the variation of the action integral (\ref{action_int}) with respect to the boundary metric $\gamma_{ab}$. To proceed further one writes the boundary metric in ADM form:
\begin{equation}
\gamma_{ab}dx^{a}dx^{b}=-N^2dt^2+\sigma_{ij}\left(d\varphi^i+V^idt\right)\left(d\varphi^j+V^jdt\right)
\end{equation}
and here the coordinates $\varphi^i$ parametrize the boundary hypersurface of constant $r$ and $N$, $V^i$ are lapse and shift correspondingly. As a result the quasilocal angular momentum can be represented in the following form \cite{Brown_PRD93, Sheykhi_PRD08}:
\begin{equation}\label{mom_int}
J=\frac{1}{8\pi}\int_{\cal B}d^{n-1}\chi\sqrt{\sigma}T_{ab}n^{a}\xi^b
\end{equation}
It is worth being noted that the latter integral is taken over a spacelike hypersurface ${\cal B}$  of the boundary $\partial{\cal M}$. And in the integral (\ref{mom_int}) we have the following notations: $\sigma$ is the determinant of the metric $\sigma_{ij}$, $n^a$ is the unit normal on the boundary ${\cal B}$ and $\xi^a$ is the rotational Killing vector $(\xi^{a}=\partial/\partial\varphi)$ . The hypersurface ${\cal B}$ is taken in such a way that it contains the orbits of the rotational Killing vector $\xi^a$.  Having used the relation (\ref{mom_int}) we obtain the evident form of the angular momentum:
\begin{equation}\label{ang_mom}
J=\frac{(n+\alpha^2-2)(n-\alpha^2)}{8\pi n(n-2)(1+\alpha^2)^2}b^{2(n-2)\gamma}\omega_{n-1}am
\end{equation}
We note that the obtained relation for the angular momentum coincides with the corresponding relation calculated in \cite{Sheykhi_PRD08} where similar slowly rotating black hole but without the cosmological constant $\Lambda$ was examined. So, we conclude that in linear approximation over the parameter $a$ the angular momentum of the black hole (\ref{ang_mom}) does not have any imprint of the cosmological constant $\Lambda$. This conclusion is in perfect agreement with the situation for the slowly rotating Kerr-Newmann-AdS black hole where angular momentum does not depend on $\Lambda$ in case when one restricts oneself by the linear approximation over the parameter $a$. 

\section{Thermodynamics of the black hole}
In this section thermodynamics of the given above black hole will be considered. To obtain the temperature we use standard definition of the black hole\rq{}s surface gravity \cite{Bardeen_CMP73}:
\begin{equation}
\kappa^2=-\frac{1}{2}\nabla_a\xi_b\nabla^a\xi^b
\end{equation}
where $\xi^{\mu}$ is a Killing vector which is null on the horizon. Since the black hole is considered in linear approximation over the rotation parameter $a$ it can be shown easily that the black hole\rq{}s temperature takes the following form:
\begin{equation}\label{temp_0}
T=\frac{\kappa}{2\pi}=\frac{W\rq{}(r_+)}{4\pi}=-\frac{(\al^2+n-2)}{4\pi(1+\al^2)}mr_+^{(1-n)(1-\gamma)}+\frac{2(n-2)(1+\al^2)}{4\pi(1-\al^2)}b^{-2\gamma}r_+^{2\gamma-1} -\frac{2\Lambda(1+\alpha^2)}{4\pi(n-\al^2)}b^{2\gamma}r_+^{1-2\gamma}
\end{equation}
and here $r_+$ is the radius of black hole\rq{}s  horizon. It is worth being noted that here we have used the relation for the mass parameter (\ref{mass_param}) and expressed the charge parameter $q$ in terms of the mass parameter  $m$, horizon radius $r_+$ and the cosmological constant $\Lambda$ with the following substitution of the obtained relation for the charge parameter into the relation for the temperature. We also point out that given above expression for the temperature coincides with the corresponding expression obtained for the static solution \cite{Sheykhi_PRD07} and in the limit when $\Lambda=0$ one can arrive at the expression for temperature of a  slowly rotating black hole but without the cosmological constant term \cite{Sheykhi_PRD08}.

\begin{figure}[!]
\centerline{\includegraphics[scale=0.35,clip]{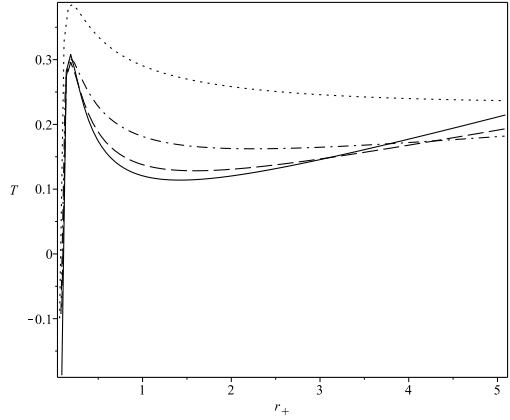}}
\caption{Temperature as a function of horizon radius $r_+$ for several different values of the dilaton coupling constant $\alpha$ in assumption that cosmological constant is fixed ($\Lambda=-1$). The correspondence of the curves is as follows: the solid curve represents the case $\al=0.1$, the dashed curve corresponds to the choice $\al=0.3$, the dash-dotted curve represents the choice $\al=0.5$ and finally the dotted curve corresponds to the  case $\al=0.7$. We also note that the other parameters in the formula (\ref{temp}) are held fixed, namely $n=3$, $b=1$ and $q=0.1$.}\label{fig_temp_1}
\end{figure}

\begin{figure}[!]
\centerline{\includegraphics[scale=0.35,clip]{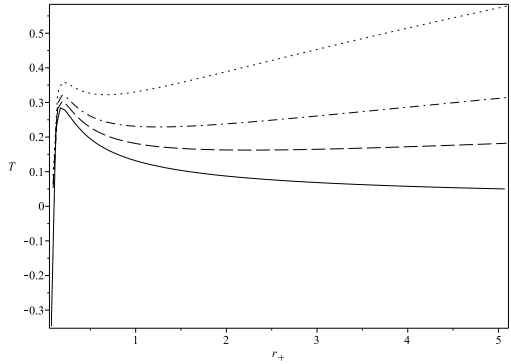}}
\caption{Temperature as a function of horizon radius $r_+$ for several different values of the cosmological constant $\Lambda$. The solid, dashed dash-dotted and dotted curves correspond to the cases $\Lambda=0$, $\Lambda=-1$, $\Lambda=-2$ and $\Lambda=-4$ respectively. The other parameters are held fixed, namely $n=3$, $\al=0.5$, $b=1$ and $q=0.1$.}\label{fig_temp_2}
\end{figure}

The mass parameter of the black hole $m$ can be expressed in terms of the horizon radius $r_+$, the cosmological constant $\Lambda$ and the charge $q$  and this representation is very convenient from the point of view of thermodynamics, because the mass of the black hole is identified with the internal energy, the horizon radius $r_+$ might be represented as a function of entropy and the parameters $q$ and $\Lambda$ also can be treated as thermodynamic values.  Now we rewrite the latter relation for the temperature taking into account the described above remark and as a result we arrive at the relation:
\begin{equation}\label{temp}
T=\frac{(1+\al^2)}{4\pi}\left(\frac{n-2}{1-\al^2}b^{-2\gamma}r^{2\gamma-1}_+-\frac{\Lambda}{n-1}b^{2\gamma}r_+^{1-2\gamma}-\frac{2q^2}{n-1}b^{2(2-n)\gamma}r_+^{2(2-n)(1-\gamma)-1}\right).
\end{equation}
We note that the written above relation for the temperature (\ref{temp}) follows the spirit of thermodynamics better than the given before relation (\ref{temp_0}), because the black hole entropy, we are going to obtain below, would be a monotonous function of the horizon radius and it means that an inverse relation where the horizon radius is a function of the entropy is well defined and two other parameters $q$ and $\Lambda$ as it has already been mentioned can be considered as thermodynamic values.  The obtained dependence of the temperature $T$ from the radius of horizon $r_+$ is rather complicated. To make it more transparent we represent it graphically for several different values of the coupling constant $\al$ and the cosmological constant $\Lambda$. The Figure [\ref{fig_temp_1}] shows the multiplicity of behaviour of the temperature for different values of the parameter $\al$. But nevertheless, there are several common features  shared by all the variants represented here. Firstly, for some small value of the horizon radius $r_+$ the temperature becomes negative and it means that the black hole of a radius below this crossing point would not exist. Secondly, here we have two extrema, specifically points of minimal and maximal temperature. The maximal temperature is higher when the parameter $\al$ is greater and the minimum of  temperature becomes lower when this parameter decreases. We also note that when the horizon radius becomes big enough the temperature also increases but the temperature goes up slowly for bigger parameter $\al$. Going a bit further we remark that the extrema points are the points of discontinuity of heat capacity which would be investigated below.
The second graph (Fig. [\ref{fig_temp_2}]) shows the $T-r_+$ dependence for several values of the cosmological constant $\Lambda$ when parameter $\al$ is kept constant. Looking at the Fig. [\ref{fig_temp_2}] one can conclude that the increase (in absolute values) of the cosmological constant gives rise to the increase of the black hole temperature. The second important point is the fact that when the cosmological constant is large enough (in absolute values) those extrema points that we have mentioned above disappear and it would be the only point which changes the character of growth of the temperature, namely from the fast growth to the regime of moderate increase.

To consider thermodynamics we should have entropy of the black hole. The entropy of the black hole is defined as a quarter of its horizon area:
\begin{equation}\label{entropy}
S=\frac{\omega_{n-1}}{4}b^{\gamma(n-1)}r^{(n-1)(1-\gamma)}_+
\end{equation}
Using the relations (\ref{mass_bh}) and (\ref{charge})  it can be easily verified that:
\begin{equation}
dM=TdS+\Phi dQ
\end{equation}
and here:
\begin{equation}
T=\left(\frac{\partial M}{\partial S}\right)_Q, \quad \Phi=\left(\frac{\partial M}{\partial Q}\right)_S
\end{equation}
are the temperature and electric potential correspondingly.

\begin{figure}
\centerline{\includegraphics[scale=0.3,clip]{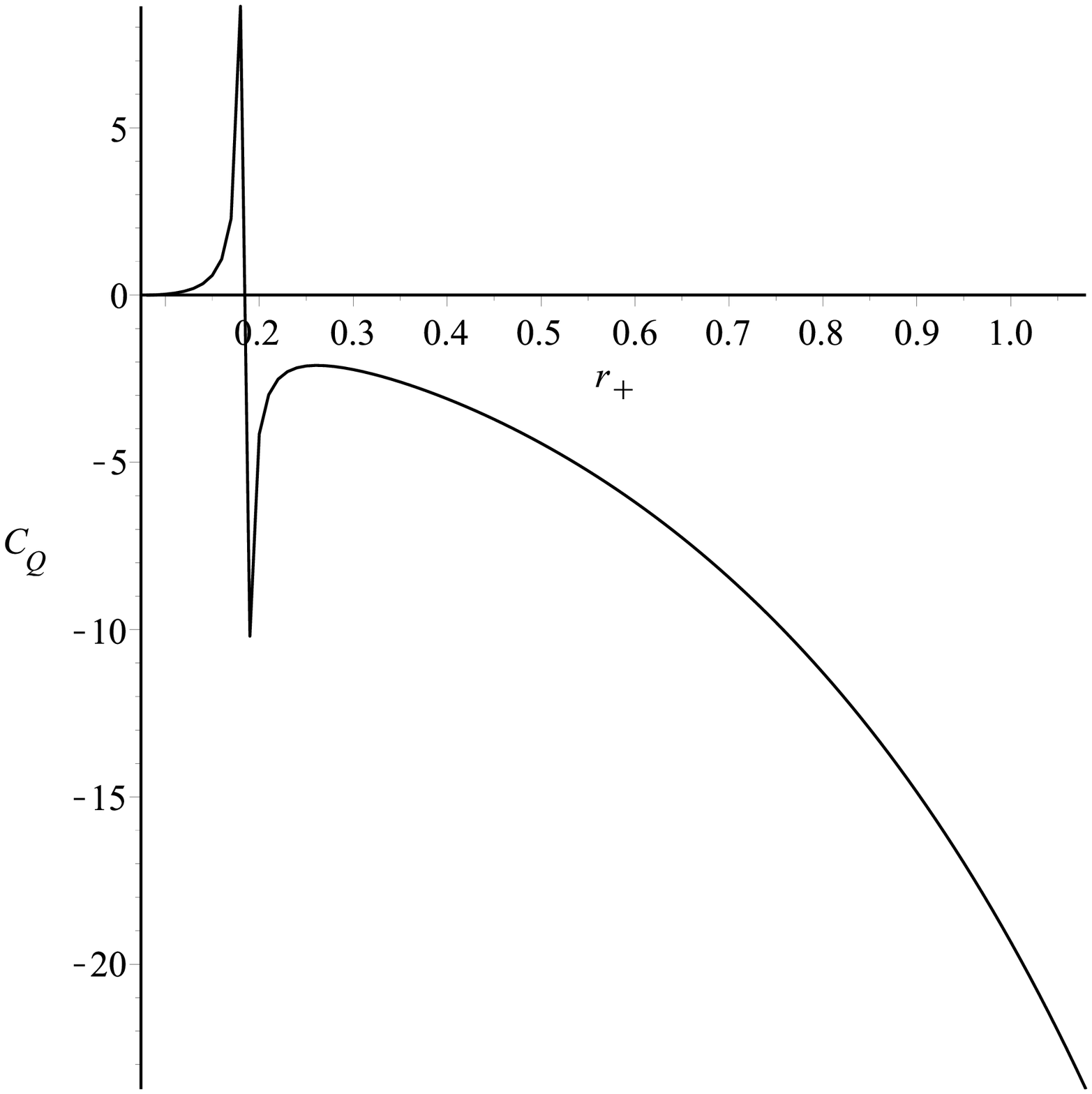}\includegraphics[scale=0.3,clip]{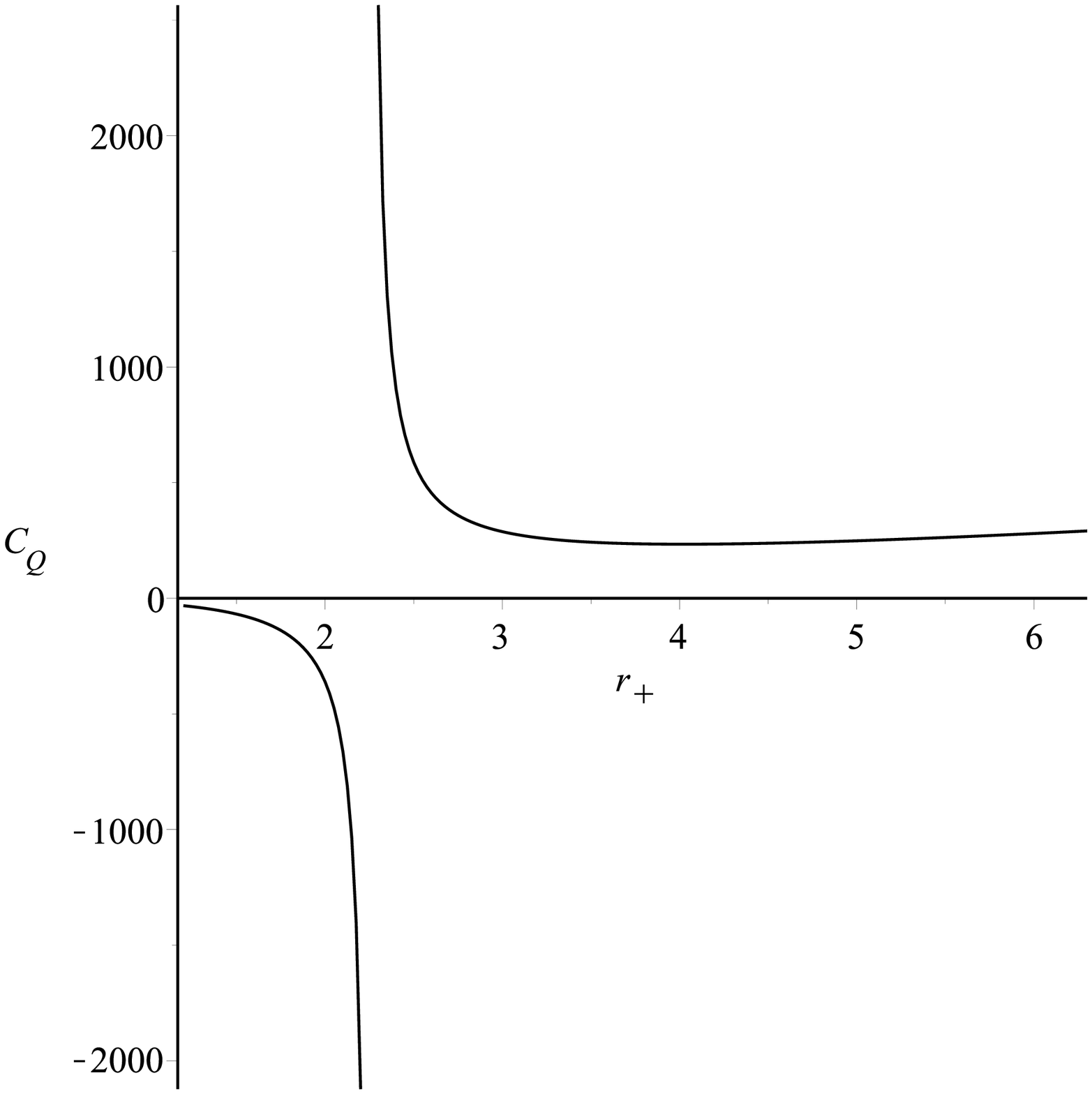}}
\caption{Heat capacity $C_Q$ as a function of horizon radius $r_+$ when the other parameters are fixed ($n=3$, $\al=0.5$, $b=1$, $q=0.1$ and $\Lambda=-1$). The left graph represents the heat capacity near the first point of discontinuity and the right one shows the heat capacity near the second discontinuity point.}\label{fig_cq}
\end{figure}

To investigate thermal stability heat capacity should be calculated:
\begin{equation}\label{heat_capacity}
C_Q=T\left(\frac{\partial S}{\partial r_+}\right)_Q\left(\frac{\partial T}{\partial r_+}\right)^{-1}_Q
\end{equation}
Having substituted  the expression for temperature (\ref{temp}) into  the latter relation  and performing  partial differentiation we obtain:
\begin{eqnarray}\label{c_q}
\nonumber C_Q=\frac{(n-1)\omega_{n-1}}{4}b^{\gamma(n-1)}r^{(n-1)(1-\gamma)-1}_+\left(\frac{n-2}{1-\al^2}b^{-2\gamma}r^{2\gamma-1}_+-\frac{\Lambda}{n-1}b^{2\gamma}r_+^{1-2\gamma}-\frac{2q^2}{n-1}b^{2(2-n)\gamma}r_+^{2(2-n)(1-\gamma)-1}\right)\\\times\left(-(n-2)b^{-2\gamma}r^{2(\gamma-1)}_+-\frac{\Lambda(1-\alpha^2)}{n-1}b^{2\gamma}r_+^{-2\gamma}-\frac{2q^2(3-2n-\alpha^2)}{n-1}b^{2(2-n)\gamma}r_+^{2((2-n)(1-\gamma)-1)}\right)^{-1}
\end{eqnarray}
As we have mentioned before in those cases when the temperature (\ref{temp}) has two extrema (namely minimal and maximal temperature) the heat capacity (\ref{c_q}) possesses the points of discontinuity for the same values of the horizon radius $r_+$. So, these points can be treated as the very same points that allow us to separate stable and unstable configurations. To comprehend the behaviour of the heat capacity $C_Q$ better we give plots of the heat capacity as a function of horizon radius $r_+$ (see the Fig.[\ref{fig_cq}]). As one can see, between the points of discontinuity the heat capacity is negative and it tells us that the system is unstable whereas for the domains with positive heat capacity the system is stable. The discontinuity of heat capacity means that the system has the phase transition of the second order. We also note, that in the domain lying left to the first discontinuity point we have the point when the heat capacity becomes equal to zero with following negative values and as it is easy to conclude that it is the very same point when the temperature (\ref{temp}) is equal to zero. As it has been mentioned before this point is the boundary point, for horizon values $r_+$ less than this boundary the temperature (\ref{temp}) would be negative and the system do not exist.

\begin{figure}
\centerline{\includegraphics[scale=0.3,clip]{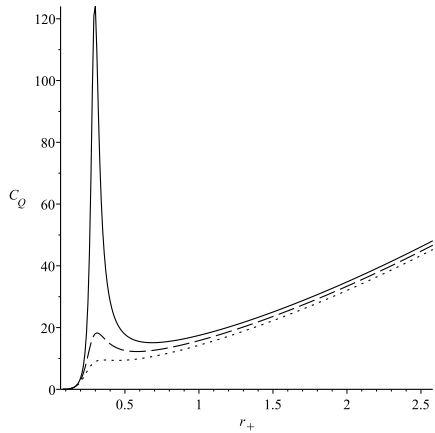}}
\caption{Heat capacity $C_Q$ as a function of horizon radius $r_+$ when discontinuity disappears ($n=3$, $\al=0.5$, $b=1$, $q=0.1$). The solid curve corresponds to the value $\Lambda=-7.4$, dashed curve corresponds to $\Lambda=-8.5$ and finally the dotted line is related to $\Lambda=-10$.}\label{fig_cq3}
\end{figure}

When the cosmological constant $\Lambda$ exceeds in absolute value some critical one, namely when the two extrema disappear, the heat capacity would not have discontinuity. The behaviour of the heat capacity in this case is shown on the Figure [\ref{fig_cq3}]. We can conclude that when the  cosmological constant goes up in absolute value the points of discontinuity become closer to each other and for some specific (critical) value of $\Lambda$ they merge and the following increase of $\Lambda$ gives rise to transformation of the merged discontinuity into a peak of a finite height. 
The further increase of the absolute value of $\Lambda$ leads to decrease of the peak of heat capacity with its following disappearance. So, we can conclude that when the absolute value of $\Lambda$ is large enough the heat capacity $C_Q$ is positive for all values of $r_+$ greater than mentioned above boundary value when the temperature $T$ (\ref{temp}) equals to zero and as a result the system is thermodynamically stable for all these values of $r_+$.

\section{Extended thermodynamics}
For long time it has been supposed that black hole thermodynamics should be considered in a \lq\lq{}fixed\rq\rq{} background, which means that for the theories with cosmological constant, namely for AdS-type black holes the cosmological constant is held fixed. An extension of standard thermodynamic phase space has been proposed recently.  It was supposed that cosmological constant might be varied \cite{Kastor_CQG09}. It leads to interesting  physical consequences and implications. As it has been emphasized earlier the extended thermodynamic phase space might bring better understanding of the black hole thermodynamics from the broader point of view and it allows to reveal new ties with the thermodynamics of real systems, namely liquid-gas systems which are described by the Van der Waals equation of state \cite{Kubiznak_JHEP12, Kubiznak_JHEP12_2}. It was also noted that the variation of cosmological constant allows to solve some important problems in the black hole thermodynamics, namely in the extended phase space Smarr relation can be recovered, it was also pointed out that the extended thermodynamics which introduces the notion of thermodynamic volume of a black hole which satisfies the inverse isoperimetric inequality \cite{Cvetic_PRD11}.

To develop the extended thermodynamics it was supposed that cosmological constant defines pressure due to the relation:
\begin{equation}
P=-\frac{\Lambda}{8\pi}.
\end{equation}
It was noted that matter fields which give rise to other black hole solutions might modify the latter relation. In the presence of a dilaton field the relation for pressure takes the following form: 
\begin{equation}\label{press}
P=-\frac{\Lambda}{16\pi}\left(\frac{b}{r}\right)^{2\gamma}
\end{equation}
It was pointed out that in case of extended thermodynamics the black hole\rq{}s mass should not be identified with thermodynamic internal energy but rather with the enthalpy $M=H$. Having used well known relation for the enthalpy one might obtain thermodynamic volume:
\begin{equation}
V=\left(\frac{\partial H}{\partial P}\right)_T
\end{equation}
Utilizing the latter relation we obtain the thermodynamic volume  which takes the form:
\begin{equation}\label{td_vol}
V=\frac{\omega_{n-1}(1+\alpha^2)}{n-\alpha^2}b^{(n-1)\gamma}r_+^{(n-1)(1-\gamma)+1}.
\end{equation} 
The thermodynamic volume is supposed to be positive and  from the given above relation a restriction on the parameter $\alpha$ can be obtained, namely $\alpha<\sqrt{n}$. The restriction imposed on the parameter $\al$ can be explained by the fact that when $\al>\sqrt{n}$ the third term in the function (\ref{func_w}) changes the sign and as a consequence it changes the asymptotic behaviour of the metric for large values of radius $r$ and as a result it affects drastically on the thermodynamics. It is easy to verify that in the limit $\alpha=0$ the volume of a ball with radius $r=r_+$ in $n$-dimensional space is recovered (which is supposed to be a geometrical volume of a slowly rotating charged black hole):
\begin{equation}
V_{\alpha=0}=\frac{\omega_{n-1}}{n}r_+^{n}.
\end{equation}
We also point out that both given definitions of the pressure can be used here, they change the resulting expression for the thermodynamic volume but it does not affect on the conclusions which will be derived from the following analysis.
 
The equation of state for the black hole can be obtained due to the relations (\ref{temp}), (\ref{press}) and (\ref{td_vol}) and takes the following  form:
\begin{equation}\label{eq_of_state}
P=\frac{(n-1)}{4(1+\alpha^2)}\frac{T}{r_+}-\frac{(n-1)}{16\pi}\left(\frac{n-2}{1-\alpha^2}b^{-2\gamma}r_+^{2(\gamma-1)}-\frac{2q^2}{n-1}b^{2(2-n)\gamma}r^{2((2-n)(1-\gamma)-1)}_{+}\right)
\end{equation}
To obtain \lq\lq{}physical\rq\rq{} equation of  state a transformation from the \lq\lq{}geometrical\rq\rq{} values of pressure and temperature  are usually replaced by \lq\lq{}physical\rq\rq{} ones:
\begin{equation}
[P]=\frac{\hbar c}{l^{n-1}_{Pl}}P, \quad [T]=\frac{\hbar c}{k}T,
\end{equation}
and here $l_{Pl}$ is the Planck length for $n$-dimensional case. To make the analogy with Van der Waals equation more transparent in the latter equation of state the horizon radius times $l^{n-1}_{Pl}$  is replaced by some specific volume.  

In the following  we will use geometric units $P$, $T$ but instead of the horizon radius $r_+$ we introduce specific ``volume'' $v$ in the following manner:
\begin{equation}\label{sp_vol}
v=\frac{4(1+\alpha^2)}{n-1}r_+
\end{equation}
Having used the introduced above parameter we rewrite  the equation  of state (\ref{eq_of_state})  in the form:
\begin{equation}\label{eq_of_st_2}
P=\frac{T}{v}-\frac{(n-2)(1+\alpha^2)}{4\pi(1-\alpha^2)}\kappa^{2\gamma-1}b^{-2\gamma}v^{2(\gamma-1)}+\frac{q^2}{8\pi}\kappa^{2((2-n)(1-\gamma)-1)}b^{2(2-n)\gamma}v^{2((2-n)(1-\gamma)-1)}
\end{equation}
and here $\kappa=(n-1)/(4(1+\alpha^2))$. 

It is easy to verify that the equation of state (\ref{eq_of_st_2}) has an inflection point for the temperatures below some critical value $T_c$. In order to examine this critical behaviour further the standard relations for the inflection point are used:
\begin{equation}
\left(\frac{\partial P}{\partial v}\right)_T=0, \quad \left(\frac{\partial^2 P}{\partial v^2}\right)_T=0
\end{equation}
Having used the latter relations we obtain critical specific volume $v_c$ and critical temperature $T_c$: 
\begin{equation}\label{v_c}
v_c=\left(\frac{2q^2(n+\alpha^2-1)(2n+\alpha^2-3)}{(n-1)(n-2)}\kappa^{2((2-n)(1-\gamma)-\gamma)}b^{2(3-n)\gamma}\right)^{1/(2(\gamma-(2-n)(1-\gamma)))}
\end{equation}
\begin{equation}\label{t_c}
T_c=\frac{(n-2)(n+\alpha^2-2)}{\pi(1-\alpha^2)(2n+\alpha^2-3)}\kappa^{2\gamma-1}b^{-2\gamma}v_c^{2\gamma-1}
\end{equation}
and for the critical pressure one arrives at the relation:
\begin{equation}\label{p_c}
P_c=\frac{(n-2)(n+\alpha^2-2)}{4\pi(n+\alpha^2-1)}\kappa^{2\gamma-1}b^{-2\gamma}v_c^{2(\gamma-1)}
\end{equation}
Having utilized the written above relations we obtain critical ratio in the form:
\begin{equation}\label{cr_rat}
\rho_c=\frac{P_cv_c}{T_c}=\frac{(1-\alpha^2)(2n+\alpha^2-3)}{4(n+\alpha^2-1)}
\end{equation}
It is worth emphasizing that the critical value we have obtained depends on two parameters only, namely on the dimension of space (or space-time) and dilaton parameter $\alpha$ which is given in the general action integral (\ref{action_int}). We can conclude that the critical ratio represents some universal behaviour which does not depend on the particular solution of the equations of motion obtained from the action (\ref{action_int}). We also emphasize that the obtained above relation for the critical ratio (\ref{cr_rat}) coincides with corresponding relation for static Einstein-Maxwell-dilaton black hole given in the paper \cite{Deghani_2015}. It should be noted that in order  to have conventional  thermodynamic behaviour we have some restrictions on the dilaton parameter, namely from  the latter relation for critical temperature and critical ratio we obtain $\alpha^2<1$.
 
When the dilaton parameter $\alpha=0$ the well-known critical ratio for Van der Waals gas in $n+1$-dimensional space is recovered:
\begin{equation}
\left.\frac{P_cv_c}{T_c}\right|_{\alpha=0}=\frac{2n-3}{4(n-1)}.
\end{equation}

\begin{figure}
\centerline{\includegraphics[scale=0.35,clip]{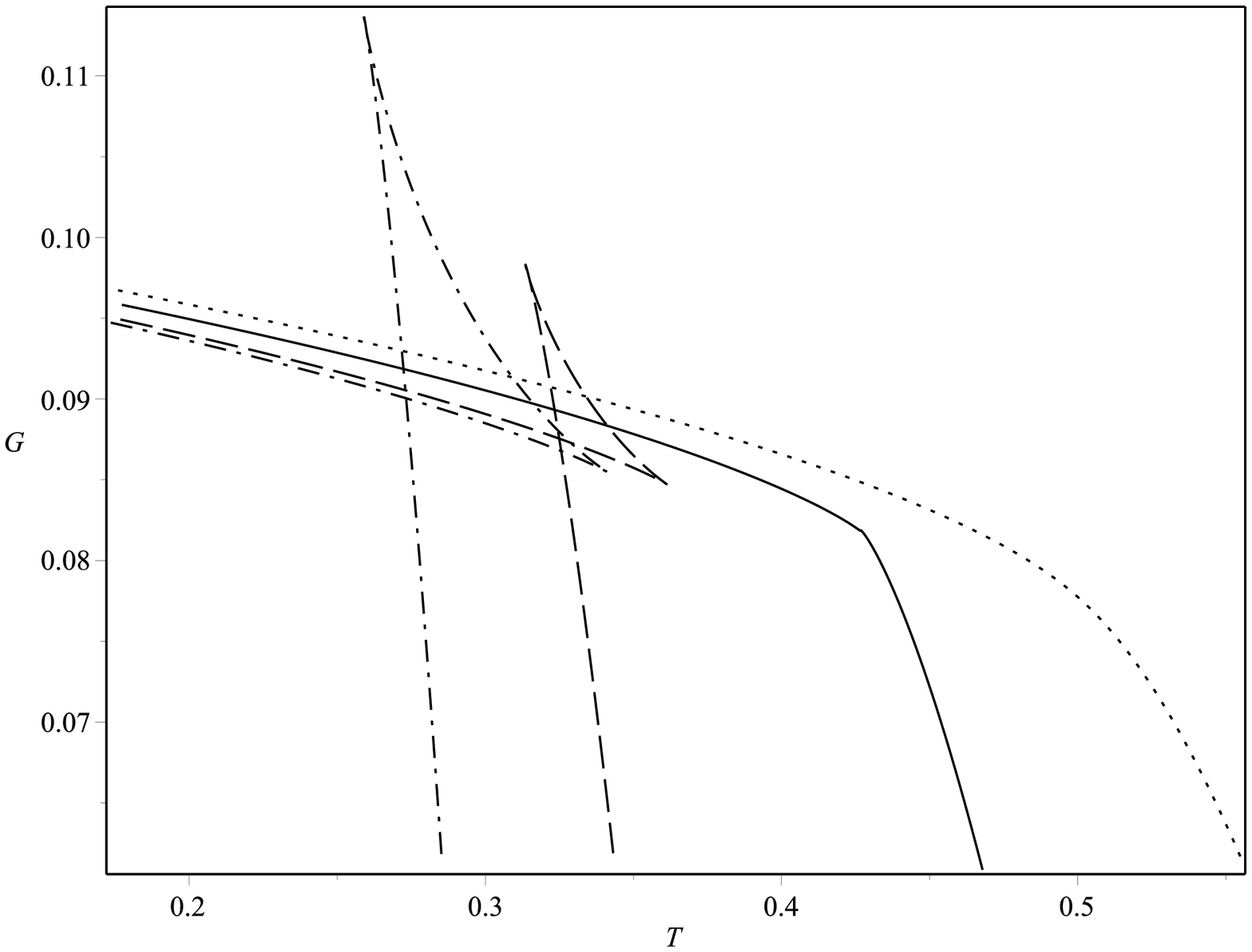}\includegraphics[scale=0.355,clip]{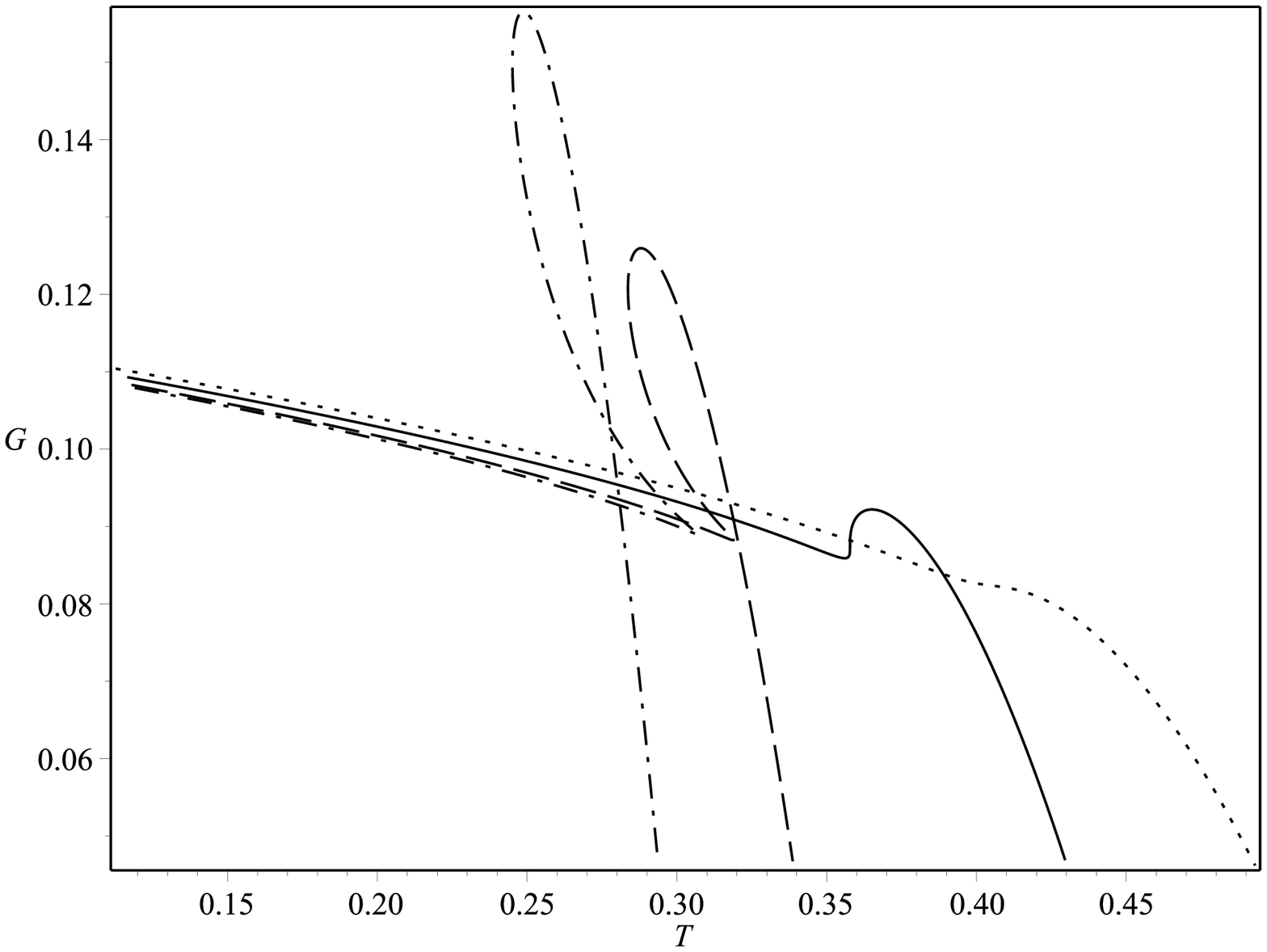}}
\caption{Gibbs free energy as a function of temperature for different values of pressure for $n=3$, $\alpha=0.1$ (left graph), $\alpha=0.5$ (right graph), $b=1$, $q=0.1$. Solid curve corresponds to the value $P=P_c$, dashed curve corresponds to the value $P=P_c/2$, dash-dotted line corresponds to the value $P=P_c/3$ and finally dotted curve corresponds to the pressure $P=3P_c/2$.}\label{fig_gibbs}
\end{figure}

As it has been mentioned above in the extended approach to thermodynamics the black hole\rq{}s mass is identified with enthalpy, so one can perform a Legendre transformation for this thermodynamic function and obtain Gibbs  free energy which is more suitable for analyzing of thermodynamic properties of systems with inherent critical behaviour. The Gibbs free energy  reads:
\begin{eqnarray}\label{Gibbs_pot}
\nonumber G(T,P)=M-TS=\frac{\omega_{n-1}(1+\alpha^2)b^{(n-1)\gamma}}{16\pi}\left(\frac{n-2}{\alpha^2+n-2}b^{-2\gamma}r_+^{(n-1)(1-\gamma)+2\gamma-1}-\right.\\\left.\frac{16\pi(1-\alpha^2)P}{(n-1)(n-\alpha^2)}r_+^{(n-1)(1-\gamma)+1}+\frac{2q^2(\alpha^2+2n-3)}{(n-1)(\alpha^2+n-2)}b^{2(2-n)\gamma}r_+^{(3-n)(1-\gamma)-1}\right)
\end{eqnarray}  
It is worth emphasizing that the Gibbs free energy is a function of temperature $T$ and pressure $P$ so in the latter relation parameter $r_+$  should be treated as a function of $T$ and $P$ through the equation of state (\ref{eq_of_state}). In the limit $\alpha=0$ the written above relation leads to the Gibbs free energy for $n+1$-dimensional RN-AdS black hole \cite{Kubiznak_JHEP12_2}
\begin{equation}
G(T,P)=\frac{\omega_{n-1}}{16\pi}\left(r_+^{n-2}-\frac{16\pi P}{n(n-1)}r_+^n+\frac{2q^2(2n-3)}{(n-1)(n-2)}r_+^{2-n}\right).
\end{equation}
The behaviour of the Gibbs free energy (\ref{Gibbs_pot}) as function of temperature for fixed values of pressure is represented on the Fig. [\ref{fig_gibbs}]. One can see that for small values of dilaton parameter $\alpha$  the Gibbs free energy has swallowtail behaviour (in our case $\alpha=0.1$), which is typical for Van der Waals system as well as for RN-AdS black hole \cite{Kubiznak_JHEP12_2}. This swallowtail behaviour of the Gibbs free energy means that the system has a phase transition of the first order. 
For bigger values of the dilaton parameter (for example, to make this new effect visible, we have chosen $\alpha=0.5$) the behaviour of the Gibbs potential is qualitatively different than for RN-AdS black hole, namely for the critical isobar ($P=P_c$) a specific maximum appears. For the pressures below the critical one a unique loop forming domain appears which does not happen in the case without dilaton parameter. Then for some isobar ($P_l<P_c$) a closed loop is formed and for the pressures below that new characteristic value $P<P_l$  one has a self -intersecting part of graph which consists of a swallowtail piece followed by some closed loop. Similarly to other systems where thermodynamics is described by a Van der Waals-like equation of state for the isotherms (or isobars) below the critical one we have a domain of instability which is characterized by a negative value of isothermal compressibility $\kappa_{T}$.  The general form of the isothermal compressibility is as follows:
\begin{equation} 
\kappa_T=-\frac{1}{V}\left(\frac{\partial V}{\partial P}\right)_T.
\end{equation}
For our equation of state (\ref{eq_of_st_2}) we can write:
\begin{equation}\label{isot_comp}
\kappa_T=\frac{n+\alpha^2}{1+\alpha^2}\left(P-\frac{n-2}{4\pi}\kappa^{2\gamma-1}b^{-2\gamma}v^{2(\gamma-1)}+\frac{q^2}{8\pi}\frac{2n+\alpha^2-3}{1+\alpha^2}\kappa^{2((2-n)(1-\gamma)-1)}b^{2(2-n)\gamma}v^{2((2-n)(1-\gamma)-1)}\right)^{-1}
\end{equation}
It is known that  the isothermal compressibility (\ref{isot_comp}) should be considered as a function of temperature $T$ and pressure $P$. To do so one should use the equation of state (\ref{eq_of_st_2}) and express volume $v$ in terms of mentioned above thermodynamic variables, but due to complicated form of the equation (\ref{eq_of_st_2}) it is impossible to perform in general case.  It can be easily checked that the latter equation (\ref{isot_comp}) gives rise to the conclusion that for the isobars above the critical one ($P>P_c$) the parameter $\kappa_T$ is always positive which means that the system is stable. Similarly one can talk about thermodynamic stability of the system for all the isotherms above the critical one ($T>T_c$). At the critical point ($P=P_c$, $T=T_c$) we have standard phase transition of the second order which is typical for other Van der Waals systems at the critical point.

\begin{figure}[!]
\centerline{\includegraphics[scale=0.35,clip]{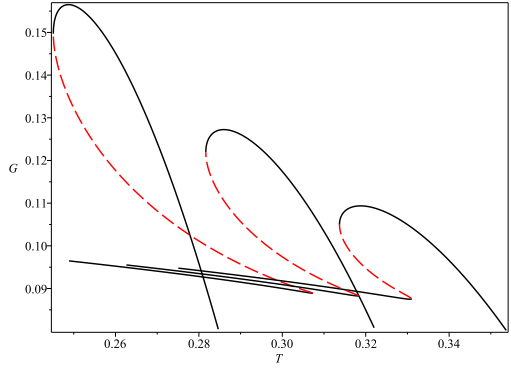}}
\caption{Gibbs free energy as a function of temperature for several values of pressure for $n=3$, $\alpha=0.5$, $b=1$, $q=0.1$ which demonstrates instability region. Instability domains are represented by red dashed parts of the curves}\label{gibbs_inst}
\end{figure}

For isobars in the region $P_l<P<P_c$ when loop is forming we have a domain of instability ($\kappa_T<0$) but because the loop is not formed yet we have a domain of discontinuity of the Gibbs free energy. Quite recently it has been pointed out that due to the discontinuity of the Gibbs free energy one has a novel phase transition of the zeroth order in that region \cite{Dehyadegari_2017}. For the isobars below the characteristic one ($P<P_l$) when the loop is formed the phase transition becomes of the first order without discontinuity of the Gibbs free energy. To make this fact more transparent let us look at the Fig. [\ref{gibbs_inst}]. It shows the transformation from the phase transition of the zeroth order into the phase transition of the first one when the pressure is decreasing. The right curve corresponds to the pressure in the region ($P_l<P<P_c$) when a closed loop has not formed yet and dashed part of the curve represent the instable domain. The middle curve corresponds to the specific pressure $P_l$ when closed loop has just formed and the zeroth order phase transition turns into the first order phase transition. The left curve corresponds to the pressure $P<P_l$ with a swallowtail part and an additional loop and in this case we have typical first order phase transition.

\begin{figure}[!]
\centerline{\includegraphics[scale=0.35,clip]{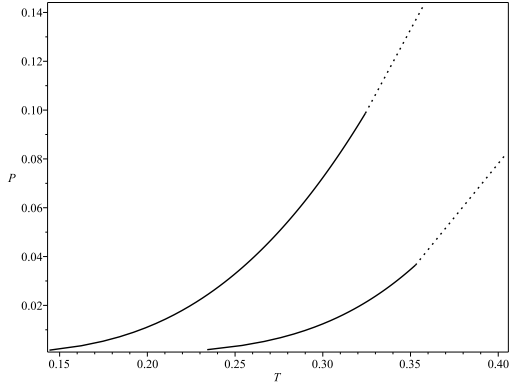}}
\caption{$P-T$-diagram (coexistence curve) for two phases for $n=3$, $\alpha=0.5$ (the left graph) or $\alpha=0.7$ (the right one), $b=1$, $q=0.1$. Solid parts of the curves represent the domains where the first order phase transition takes place and the dotted parts of the curves represent the range where the zeroth order phase transition happens.}\label{PT_curve}
\end{figure}

It is well-known that Gibbs potential is constant during the first order phase transition. This fact helps us to obtain coexistence curve for two phases. To do so we use Maxwell's equal area law. On an isotherm with the temperature $T<T_c$ two points with parameters $(P_0,v_1)$ and $(P_0,v_2)$ satisfy Maxwell's equal area law:
\begin{equation}\label{max_law}
P_0(v_2-v_1)=\int^{v_2}_{v_1}P{\rm d}v
\end{equation}
Having substituted in the right hand side of the latter relation the equation of state (\ref{eq_of_st_2}) and performing integration we obtain:
\begin{equation}\label{max_law_int}
P_0(v_2-v_1)=T\ln{\left(\frac{v_2}{v_1}\right)}+\frac{1+\alpha^2}{1-\alpha^2}A\left(v^{2\gamma-1}_2-v^{2\gamma-1}_1\right)+\frac{1+\alpha^2}{3-2n-\alpha^2}B\left(v^{2(2-n)(1-\gamma)-1}_2-v^{2(2-n)(1-\gamma)-1}_1\right)
\end{equation}
and here we denote:
\begin{equation}
A=\frac{(n-2)(1+\alpha^2)}{4\pi(1-\alpha^2)}\kappa^{2\gamma-1}b^{-2\gamma}, \quad B=\frac{q^2}{8\pi}\kappa^{2((2-n)(1-\gamma)-1)}b^{2(2-n)\gamma}
\end{equation}
From the equation of state (\ref{eq_of_st_2}) one  can obtain:
\begin{equation}\label{temp_rel}
T\left(\frac{1}{v_2}-\frac{1}{v_1}\right)-A\left(v^{2(\gamma-1)}_2-v^{2(\gamma-1)}_1\right)+B\left(v^{2((2-n)(1-\gamma)-1)}_2-v^{2((2-n)(1-\gamma)-1)}_1\right)=0
\end{equation}
To find coexistence relation we introduce new parameter $x$  as a ratio of two volumes: $x=v_1/v_2$ ($0<x<1$). Using introduced parameter we rewrite the latter relation (\ref{temp_rel}) in the following way:
\begin{equation}\label{temp_x}
T=\frac{x}{x-1}\left(Av^{2\gamma-1}_2\left(1-x^{2(\gamma-1)}\right)-Bv^{2(2-n)(1-\gamma)-1}_2\left(1-x^{2((2-n)(1-\gamma)-1)}\right)\right).
\end{equation}
Taking into account the relations (\ref{max_law_int}) and (\ref{temp_x}) we can find the relation for the volume of one phase as a function of parameter $x$:
\begin{equation}\label{v_2func}
v^{2(\gamma-(2-n)(1-\gamma))}_2=\frac{B}{A}\left(\frac{\frac{x}{x-1}\ln{x}(1-x^{2((2-n)(1-\gamma)-1)})+\frac{2(n+\al^2-1)}{3-2n-\alpha^2}\left(1-x^{2(2-n)(1-\gamma)-1}\right)}{\frac{x}{x-1}\ln{x}(1-x^{2(\gamma-1)})-\frac{2}{1-\alpha^2}\left(1-x^{2\gamma-1}\right)}\right)
\end{equation}
Having used the relation for temperature (\ref{temp_x}) we can rewrite the equation of state (\ref{eq_of_st_2}) as a function of $x$:
\begin{equation}\label{press_x}
P=\frac{1}{x-1}\left(Av^{2(\gamma-1)}_2\left(1-x^{2\gamma-1}\right)-Bv^{2((2-n)(1-\gamma)-1)}_2\left(1-x^{2(2-n)(1-\gamma)-1}\right)\right).
\end{equation}
We point out that when we set $x=1$ in the written above equations (\ref{temp_x}), (\ref{v_2func}) and (\ref{press_x}) we recover critical values for temperature $T_c$ (\ref{t_c}) volume $v_c$ (\ref{v_c}) and pressure $P_c$ (\ref{p_c}) respectively.
The latter three equations allow one to obtain the coexistence curve between two phases. It should be noted that these equations give rise to coexistence relation (P-T diagram) just for the region where the first order phase transition takes place ($P<P_l$, $T<T_l$). In the region $P_l<P<P_c$ and $T_l<T<T_c$ the coexistence curve can be approximated by a line joining the points ($T_l,P_l$) and ($T_c,P_c$). Analyzing the behaviour of the coexistence curve (the Fig.\ref{PT_curve}) one can conclude that the domain where the zeroth order phase transition transition takes place extends with increasing of the dilaton parameter $\alpha$. The ending points of these curves represent the points of the second order phase transition similarly as for other Van der Waals systems. It was pointed out \cite{Dehyadegari_2017} that for the absent dilaton parameter ($\alpha=0$) the zeroth order phase transition disappears and the coexistence curve is typical for Van der Waals systems where the first order phase transition's domain ends up at the critical point of the second order phase transition. 
The slope of the coexistence curve can be calculated with the help of the Clapeyron equation:
\begin{equation}\label{Clapeyron}
\frac{dP}{dT}=\frac{L}{T(v_2-v_1)},
\end{equation}
where  $L=T(s_2-s_1)$ is the latent heat of the phase  change and here $s_1$ and $s_2$ represent corresponding entropies for the first  and the second phases respectively. In our case these two phases represent the black holes of different sizes, namely the so called small and large black holes \cite{Kubiznak_JHEP12_2,Zhao_AHEP16,H_F_Li_2016} and the latent heat represents the gain or loss of the mass of the black hole during the phase transition \cite{Xu_IJMPD17}. From the latter equation one can obtain the relation for the latent heat of the phase change:
\begin{equation}\label{latent_heat}
L=T\frac{dP}{dT}(v_2-v_1)=v_2(1-x)T(x)\frac{dP}{dx}\frac{dx}{dT}
\end{equation}
Taking into account the relations (\ref{temp_x}), (\ref{v_2func}) and (\ref{press_x}) one can represent the latent heat of the phase change as a function of temperature $T$ but due to complicated relation that appears here we show only graphical dependence.

\begin{figure}
\centerline{\includegraphics[scale=0.333,clip]{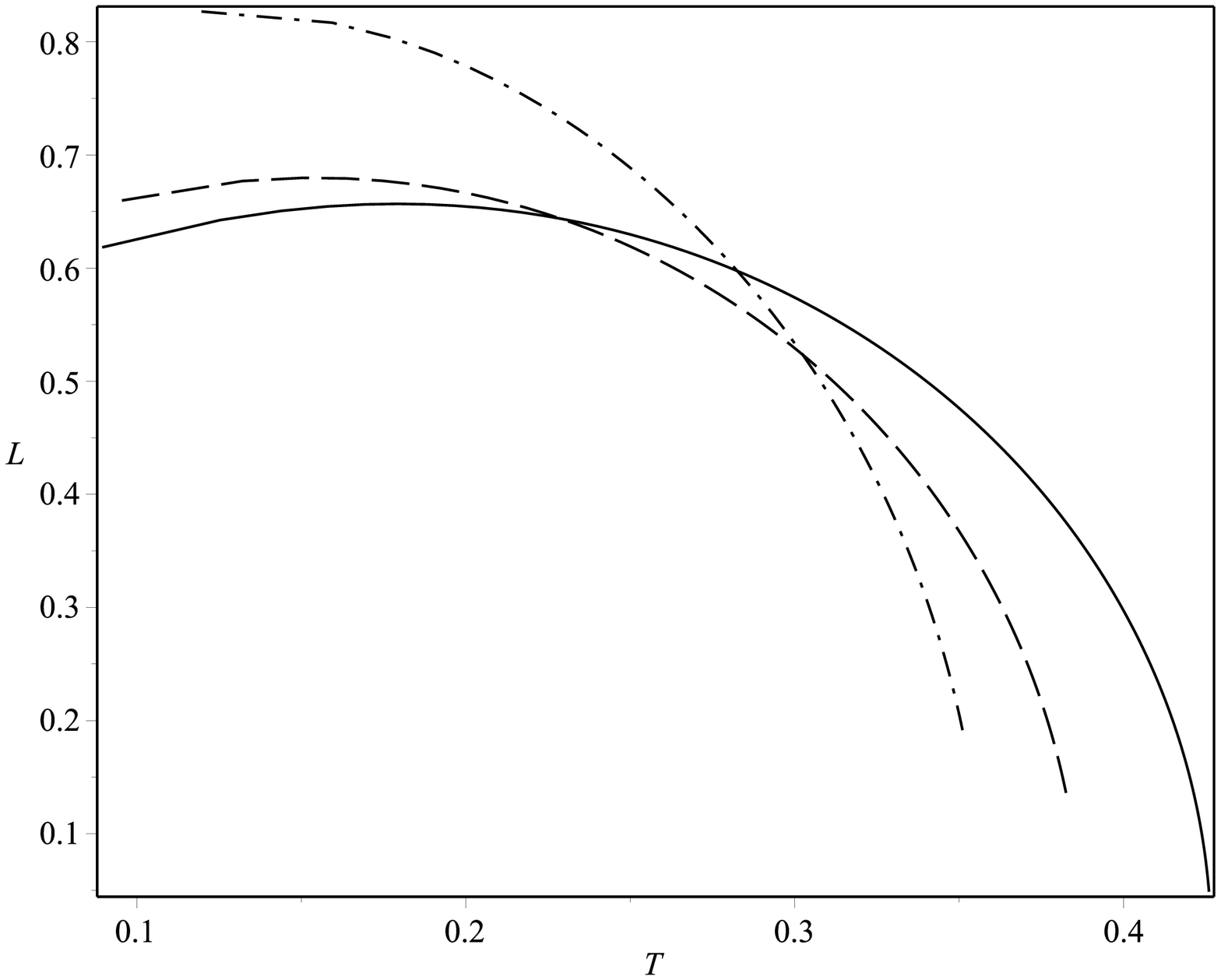}\includegraphics[scale=0.35,clip]{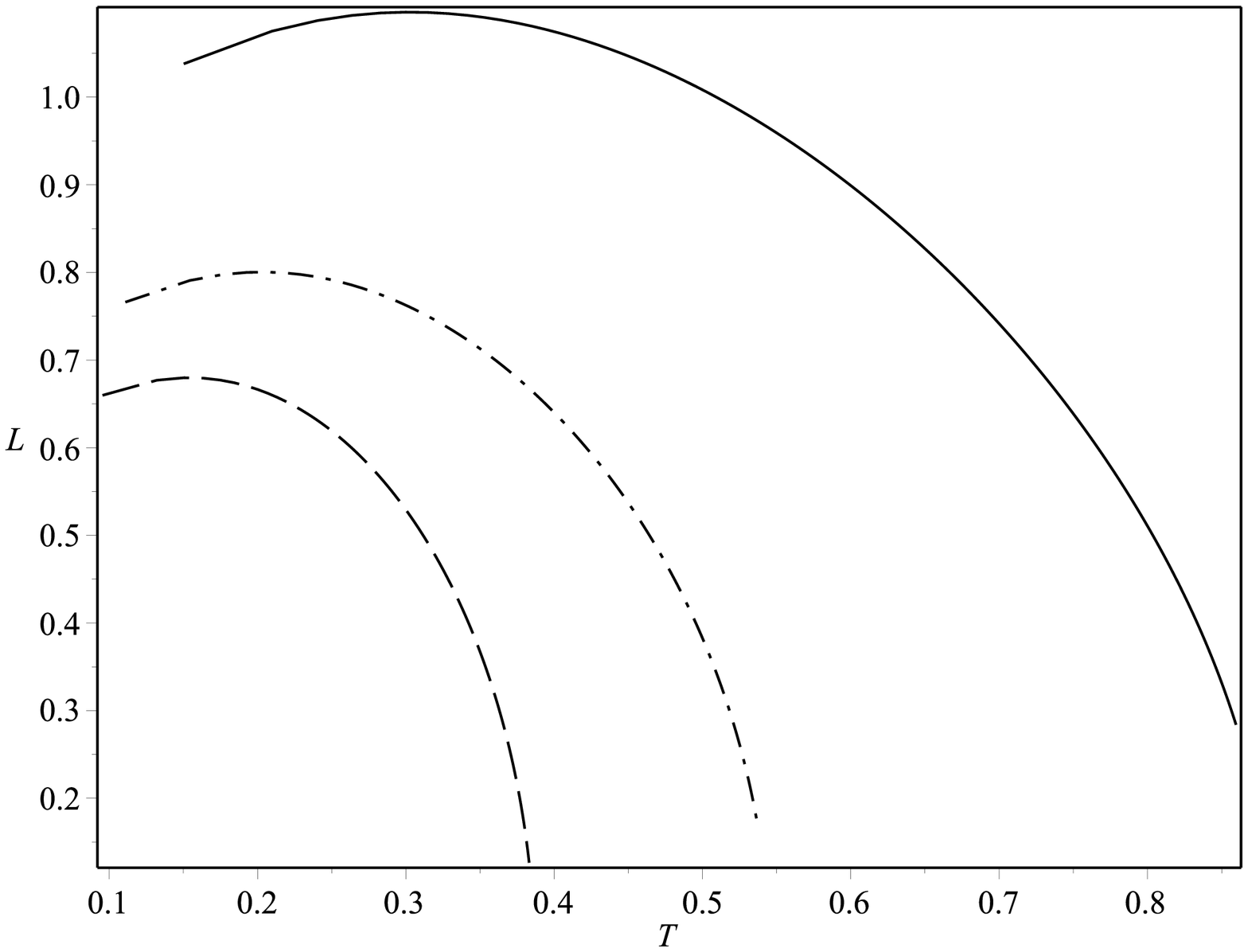}}
\caption{The left graph represents latent heat $L$ as a function of temperature $T$ for different values of $\alpha$ when $n$ is fixed, namely $n=3$, ( in particular $b=1$, $q=0.1$ for both graphs) and $\alpha=0.1$ (solid curve), $\alpha=0.3$ (dashed curve) and $\al=0.5$ (dash-dotted curve). The right graph demonstrates the $L-T$ dependence for fixed $\al=0.3$ and several values of $n$, namely $n=3$ (dashed curve), $n=4$ (dash-dotted curve), $n=6$ (solid curve).}\label{fig_lat_heat}
\end{figure}

The Figure [\ref{fig_lat_heat}] represents the two types of possible changes. The left graph shows the dependence for fixed value of $n$ and for several different values of the parameter $\al$. One can see that for smaller values of $\al$ the $L=L(T)$ dependence is nonmonotonic with some specific maximum point and than decreasing to zero. Increasing of the parameter $\al$ gives rise to the shift of the maximum point  to the lower temperatures. At the same time increasing of parameter $n$ when $\al$ is fixed leads to raising of absolute value of the latent heat $L$ with corresponding shift of the maximum point to higher temperatures. It should be pointed out that for the domain where we have the phase transition of the zeroth order it is not possible to calculate the latent heat with the help of the relation (\ref{Clapeyron}), because this formula is obtained under the assumption that the Gibbs potential is continuous which is not obeyed for the phase transition of the zeroth order. But reaching the critical temperature the phase transition transforms into the phase transition of the second order where the latent heat is equal to zero, so any of the represented curves should be continued by a sort of other curve reaching zero latent heat at the critical temperature corresponding by very same parameters as $n$, $\al$, $b$ and $q$ which represents the given curve.

\section{Critical exponents}
Since the system possesses a critical point it is interesting to investigate the behaviour of the system in the domain close to the critical point. In general we follow the approach developed in the Ref. \cite{Kubiznak_JHEP12,Kubiznak_JHEP12_2}.

To obtain critical exponent $\alpha$ one should utilize the relation for the entropy (\ref{entropy}) and represent it as a function of the temperature $T$ and thermodynamic volume $V$ (\ref{td_vol}):
\begin{equation}
S(T,V)=\frac{\omega_{n-1}}{4}b^{(n-1)\gamma}\left[\frac{(n-\alpha^2)}{\omega_{n-1}(1+\alpha^2)}b^{-(n-1)\gamma}V\right]^{(n-1)(1-\gamma)/((n-1)(1-\gamma)+1)}
\end{equation}
The written relation shows that the entropy does not depend on the temperature $T$ explicitly, thus heat capacity $C_V=0$ and as a result corresponding critical exponent $\alpha=0$ .

To find other critical exponents we rearrange the Van der Waals equation introducing reduced variables:
\begin{equation}
p=\frac{P}{P_c}, \quad \tau=\frac{T}{T_c}, \quad \nu=\frac{v}{v_c},
\end{equation} 
where $P_c$, $T_c$ and $v_c$ are corresponding critical values given by the relations (\ref{p_c}-\ref{v_c}). As a result the equation of state (\ref{eq_of_st_2}) can be represented in the form:
\begin{equation}\label{red_eq_st}
p=\frac{4(n+\alpha^2-1)}{(1-\alpha^2)(2n+\alpha^2-3)}\frac{\tau}{\nu}-\frac{(1+\alpha^2)(n+\alpha^2-1)}{(1-\alpha^2)(n+\alpha^2-2)}\nu^{2(\gamma-1)}+\frac{(1+\alpha^2)}{(n+\alpha^2-2)(2n+\alpha^2-3)}\nu^{2((2-n)(1-\gamma)-1)}
\end{equation} 
In the limit $\alpha=0$ one arrives at:
\begin{equation} 
p=\frac{4(n-1)}{(2n-3)}\frac{\tau}{\nu}-\frac{n-1}{n-2}\nu^{-2}+\frac{1}{(n-2)(2n-3)}\nu^{2(1-n)}
\end{equation}
The latter equation coincides with corresponding equation of state obtained in the Ref.\cite{Kubiznak_JHEP12_2}.  

The equation (\ref{red_eq_st}) can be rewritten in the from:
\begin{equation}
p=\frac{1}{\rho_c}\frac{\tau}{\nu}+h(\nu)
\end{equation}
where $\rho_c$ is the critical ratio given by the relation (\ref{cr_rat}) and $h(\nu)$ might be even more general than in the relation (\ref{red_eq_st}). In the neighbourhood  of the critical point ($\tau=1,\nu=1$) one can represent
\begin{equation}
\tau=1+t, \quad \nu=(1+\omega)^{1/z} 
\end{equation}
and it should be noted that $z>0$.  Taking into account the definition of the critical point one can expand the right hand side of the relation (\ref{red_eq_st}) into the series and as a consequence we arrive at the relation
\begin{equation}\label{eq_exp}
p=1+At-Bt\omega-C\omega^3+{\cal O}(t\omega^2,\omega^4)
\end{equation}
and here 
\begin{equation}
A=\frac{1}{\rho_c}, \quad B=\frac{1}{z\rho_c}, \quad C=\frac{1}{z^3}\left(\frac{1}{\rho_c}-\frac{1}{6}h^{(3)}(1)\right)
\end{equation}
We assume that $C>0$. Differentiating  the equation (\ref{eq_exp}) with respect to the $\omega$ we obtain:
\begin{equation}\label{max_law}
{\rm d}P=-P_c\left(Bt+3C\omega^2\right){\rm d}\omega
\end{equation} 
Having used Maxwell\rq{}s area law and denoting $\omega_l$ and $\omega_s$ \lq\lq{}volumes\rq\rq{} of large and small black holes respectively we obtain:
\begin{equation}
p=1+At-Bt\omega_l-C\omega^3_l=1+At-Bt\omega_s-C\omega^3_s
\end{equation} 
The latter equation gives us nontrivial solution:
\begin{equation}
\omega_s=-\omega_l=\sqrt{-\frac{Bt}{C}}
\end{equation}
As a result we obtain:
\begin{equation}
\eta=V_c\left(\omega_l-\omega_s\right)=2V_c\omega_l\sim(-t)^{1/2}
\end{equation}
From the written equation it immediately follows that:
\begin{equation}
\beta=\frac{1}{2}
\end{equation} 
To calculate the exponent $\gamma$ we use again the relation (\ref{max_law}). One can write:
\begin{equation}
\kappa_T=-\frac{1}{V}\left(\frac{\partial V}{\partial P}\right)_T\sim\frac{1}{P_cBt} , \quad \Rightarrow \quad \gamma=1.
\end{equation}
Now it follows from the ``critical isotherm'' ($t=0$) that:
\begin{equation}
p-1=-C\omega^3, \quad \Rightarrow \quad \delta=3.
\end{equation}
One can see that the obtained critical exponents are the same as for the RN-AdS black hole, so we can conclude that the presence of dilaton field giving serious influence on the thermodynamics and critical behaviour does not change the values of critical exponents.

\section{Conclusions}
In our work we have considered a slowly rotating black hole in the framework of Einstein-Maxwell-dilaton theory. In this theory an additional dilaton potential term of the so called Liouville form is taken into account and this form allows us to obtain cosmological constant term. Similar problem was examined in the works \cite{Sheykhi_PRD08, Sheykhi_PRD08_02} but in the first one of them the cosmological constant term is not  taken into account while in the second one this term takes a bit different form. We have obtained the solution of corresponding Einstein and field equations which is in a perfect agreement with the results of the works \cite{Sheykhi_PRD07,Sheykhi_PRD08, Sheykhi_PRD08_02} and in the limit when dilaton parameter goes down to zero the slowly rotating  Kerr-Newmann solution is recovered \cite{Aliev_PRD06}. It should be pointed out that the solution obtained here can be treated as a generalization of the corresponding solution in \cite{Sheykhi_PRD08}. We note that the obtained solution possesses complicated causal structure which in general aspects is analogous to nonrotating dilaton black hole \cite{Sheykhi_PRD07}.

We have investigated black hole thermodynamics in assumption that the cosmological constant is held fixed. Firstly, it was shown that for negative cosmological constant temperature has nonmonotonous behaviour as a function of horizon radius $r_+$. It is shown that for small radius $r_+$ the temperature (\ref{temp}) becomes negative and it gives rise to the conclusion that a black with such a small radius would be unstable. It is also demonstrated that the point when the temperature becomes equal to zero is the very same point when the heat capacity $C_Q$ changes its sign from positive onto a negative one and this fact can be taken as additional evidence to confirm the thermodynamic instability of the black hole.   The heat capacity was shown to have two points of discontinuity which separate stable and unstable domains. But for cosmological constant large enough in absolute value the points of discontinuity merge with each other and finally disappear transforming into high peak which diminishes with the following increasing of the cosmological constant (Fig.[\ref{fig_cq3}]). It means that when the cosmological constant is large in absolute value we have the black hole which might be stable for arbitrarily small radius of the horizon, that peculiarity, as far as we know, has not been paid attention to in the previous works.

We also examined the thermodynamics using the extended technique. It was shown that the obtained black hole solution allows one to obtain equation of state of Van der Waals type (\ref{eq_of_state}) (or (\ref{eq_of_st_2})). Similarly to recently obtained results of the work \cite{Dehyadegari_2017} our system possesses a domain of the first as well as of the zeroth order phase transitions. The appearance of the zeroth order phase transition is directly related to the existence of dilaton-Maxwell fields coupling which is described by the parameter $\al$. The main conclusion we should point out here is the fact that the stronger this coupling is, the wider domain where the zeroth order phase transition takes place becomes. For small enough $\al$ this domain, where zeroth order phase transition happens, is negligibly small and it is hardly visible on the graph (see the left graph of the Fig. [\ref{fig_gibbs}]),  but when the coupling parameter $\al$  goes up this domain drastically increases and becomes notable on the graph.  We also note that the zeroth order phase transition takes place also for other systems with dilaton-electromagnetic coupling with different type of electromagnetic field action \cite{Dayyani_2017}, but in that work it was pointed out that the discontinuity of the Gibbs potential which gives rise to the zeroth order phase transition is related not only to the dilaton-electromagnetic field coupling constant $\al$. In our case  we can state that the existence of the zeroth order phase transition is completely caused by the coupling between the fields, in case this coupling disappear the zeroth order transition would not occur. Using the equation of state and the Maxwell's equal area law we also obtained coexistence relation for the system in the domain where the first order phase transition takes place. Having used Clapeyron equation the latent heat was calculated numerically, and we note that these calculations are valid on the very same domain. Finally, we calculated critical exponents. They are shown to be the same as for Einstein-Maxwell black hole.  

\section{Acknowledgements}
This work was partly supported by Project FF-30F (No. 0116U001539) from the Ministry of Education and Science of Ukraine.

\end{document}